RESEARCH ARTICLE

# Interactive reconstructions of cranial 3D implants under MeVisLab as an alternative to commercial planning software

Jan Egger[1,2]\*, Markus Gall[1], Alois Tax[3], Muammer Üçal[3], Ulrike Zefferer[3], Xing Li[4], Gord von Campe[3], Ute Schäfer[3], Dieter Schmalstieg[1], Xiaojun Chen[4]\*

1 Institute for Computer Graphics and Vision, Faculty of Computer Science and Biomedical Engineering, Graz University of Technology, Inffeldgasse 16, Graz, Austria, 2 BioTechMed-Graz, Krenngasse 37/1, Graz, Austria, 3 Department of Neurosurgery, Medical University of Graz, Auenbruggerplatz 29, Graz, Austria, 4 Institute of Biomedical Manufacturing and Life Quality Engineering, School of Mechanical Engineering, Shanghai Jiao Tong University, China

\* egger@tugraz.at; xiaojunchen@163.com

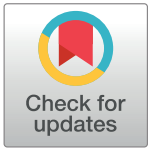

## Abstract

In this publication, the interactive planning and reconstruction of cranial 3D Implants under the medical prototyping platform MeVisLab as alternative to commercial planning software is introduced. In doing so, a MeVisLab prototype consisting of a customized data-flow network and an own C++ module was set up. As a result, the Computer-Aided Design (CAD) software prototype guides a user through the whole workflow to generate an implant. Therefore, the workflow begins with loading and mirroring the patients head for an initial curvature of the implant. Then, the user can perform an additional Laplacian smoothing, followed by a Delaunay triangulation. The result is an aesthetic looking and well-fitting 3D implant, which can be stored in a CAD file format, e.g. STereoLithography (STL), for 3D printing. The 3D printed implant can finally be used for an in-depth pre-surgical evaluation or even as a real implant for the patient. In a nutshell, our research and development shows that a customized MeVisLab software prototype can be used as an alternative to complex commercial planning software, which may also not be available in every clinic. Finally, not to conform ourselves directly to available commercial software and look for other options that might improve the workflow.

## Introduction

Cranioplasty, were a defect or deformity of the cranial bone is repaired [1], is often performed because of traumas, infections, tumors or compressions due to brain edema [2]. Even though the method is relatively safe, global intracerebral infarction can cause lethal results [3]. It is also possible, that further complications occur resulting in a loss of the cranial implant. Brommeland et al. [4] for example, showed that out of 87 patients, 37 (36%) suffered from complications, whereby even 22 lost their primary implant. The most common causes were surgical site infections, affecting eight patients (9.2%) and bone flap resorption in fourteen patients (19.7%). Furthermore, there is a broad range on different used materials to close the defect, like acrylic [5] as Data Availability Statement: All relevant data are within the paper and hosted at the public repository Figshare. Please see data hosted at Figshare at the following URL: https://figshare.com/articles/Cranial_Defect_Datasets/4659565.

Funding: The work received funding from BioTechMed-Graz in Austria (Hardware accelerated intelligent medical imaging) and the 6th Call of the Initial Funding Program from the Research & Technology House (F&T-Haus) at the Graz

University of Technology (PI: Jan Egger). Dr. Xiaojun Chen receives support by the Natural Science Foundation of China (Grant No.: 81511130089) and the Foundation of Science and Technology Commission of Shanghai Municipality (Grants No.: 14441901002, 15510722200 and 16441908400).

**Competing interests:** The authors declare no competing interests.

one of the most used ones or composite materials like hydroxyapatite-poly(methylmethacrylate) (PMMA) composites [6]. Nevertheless, using autologous bone is still considered the gold standard, but methods using osteointegration were synthetic porous implants guide the bone regeneration, call for an urgent need as well [7]. Since no cranial defect looks like the other, patient-individual implants are in general needed for a successful and precise treatment. This usually requires a careful pre-operative planning on the basis of a Computed Tomography (CT) scan of the patient's skull and the design of an exact virtual 3D model of the implant (Fig 1) [8]. However, constructing a virtual model of a cranial 3D implant is a challenging task, which often lacks of appropriate software that can be applied in the pre-planning phase. One option are commercial software products, like MIMICS, Biobuild or 3D doctor, which tend to be very complex and expensive, and hence, they are not available in every clinic. Another option is to use (free) non-medical software, like the 3D computer graphics software Blender (https://www.blender.org/), to create an implant computer model. However, due to the fact that these software tools are not intended to create medical implants and don't offer specific and convenient tools for these tasks, the design process can take up to several hours. This is mainly the case, because vertex after vertex has to be pulled into position by simple drag and drop operations. Obviously, this course of action is not a very user-friendly way to reach a satisfying result.

Other working in the area of computer-aided cranioplasty are Lee et al. [9], who present a custom implant design case for an 8-year-old boy with a large cranial defect. The raw cranial CT data of the patient was translated into the stereolithography (STL) format utilizing a computer-aided manufacturing/computer-aided design (CAM/CAD) interface software tool. Anyhow, Lee et al. do not state which CAM/CAD software tool they applied or describe it in further detail. Van der Meer and colleagues [10] present the digital planning of cranial implants for the reconstruction of skull defects. However, the overall workflow is very time-consuming and includes the usage of a generic industrial software. The software offers automated procedures and functions like "curvature-based filling" and "fill holes" that have been applied to design the final implant model. Chulvi et al. [11] demonstrate the automated generation of customized implants by linking two computer software prototypes. Thereby, a Knowledge Based System technology is the core of the model. On one side, it is able to manage and store medical data, on the other side, it is able to manage and store designer knowledge. Afterwards, this information is used for the implant design process and the research findings are based on previous studies of existing software tools, like MIMICS, 3D Slicer [12], ImLib3D, MITK, OsiriX and the Visualization Toolkit (http://www.vtk.org/). Another proposed method from Lo et al. [13] uses the mirrored healthy side of the skull as a template combined with further, time intense, optimization using the AnalyzeTM software. A CAD tool for the custom implant design for large cranial defects (>100 cm$^2$) has recently been published by Marreiros et al. [14]. The approach uses a combination of geometric morphometrics and radial basis functions for the semi-automatic implant generation. Further, the method uses symmetry and the best fitting shape to estimate missing data directly within the radiologic volume data. Finally, reconstructions of skulls like presented in [15] and [16] are not based on the mirroring technique, rather they use shape models to close the defects, which means that this work is mainly for areas where "simple" surface parts have to be reconstructed. However, according to Gilardino et al. [17], the usage of current computer-aided pre-planning systems is still quite time-consuming, but they reduce the number of following complications significantly and some softwares allow already wizard-wise packages and the creation of macro's, which can help to reduce the planning time.

In this contribution, we developed a planning prototype for 3D cranial implants within the freely available medical research platform MeVisLab (http://www.mevislab.de/) [18]. To the best of our knowledge, this is the first time cranioplasty has been introduced to the MeVisLab platform. We decided for the semi-commercial platform MeVisLab, because it is currently more stable and user friendly than the pure open source platforms, like Slicer or the two MITK toolkits (www.mitk.org/ and http://www.mitk.net/ from Germany and China). In summary, our method uses the mirrored skull as a template for generating a good fitting and aesthetic looking implant. Fitting in terms of no gaps between the bone and the implant. However, since surgeons have an interest in modifying the implants individually to a certain level, we did not design a fully-automated system performing all operations without any user interaction. Rather, the user can manually intervene in every step for specific modifications of the implant. Finally, the implant model can be stored as STL file to be used with 3D printing technology [19]. As audience for this contribution, we want primarily to target clinical/biomedical end users of our prototype. However, we also try to target researches that may want to build upon our solution. Thus, we also provide a more detailed technical description and code-sections like the paragraphs concerning the "Smoother module". We hope that the technical description makes it easier to understand the network/module and enables extensions for further features.

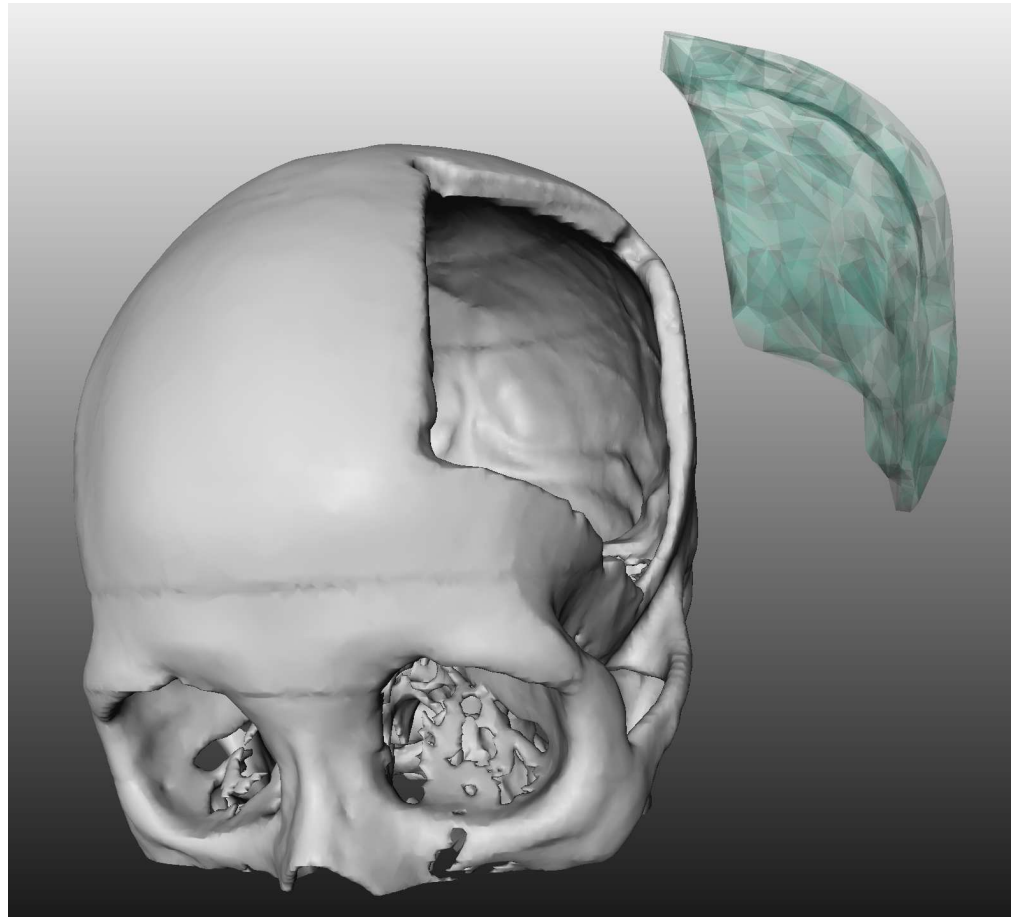

**Fig 1. 3D visualization of a patient skull with a cranial defect on the left side.** The image shows also a 3D model of a cranial implant to close the defect (light green).

This contribution is further organized as follows. Section 2 depicts details of the used material and the newly proposed integration, presents the theoretical background of the proposed mechanism and provides sufficient detail to allow the work to be reproduced. In Section 3, experimental results, including illustrations of generated implants, are presented. Section 4 gives a summary together with a short discussion to highlight the significance of the introduced achievement and lays the foundation for further work.

## Material and methods

Datasets–The datasets used for evaluating our software were stored in STL format, originally derived from CT scans with average slice thicknesses of 1 mm (the transformation from CT/DICOM data to STL can be done for example with the *WEMIsoSurface* module available under MeVisLab). However, there was no more information about any of the used files since they were only provided in STL-format from the clinical partners for anonymization purposes. In addition, the lower part, or more precisely, the mouth region, was removed from the skull to make it impossible to recognize the patients via their teeth. For testing our pre-planning tool, a variety of patient skulls, suffering from cranial defects, with variations in anatomy and location of the pathology, were made available from our clinicians. The defects ranged from small to bigger ones, like from the central, frontal bone to the dextral ethmoidal bone without affecting the orbital bow or a bigger cranial defect on the sinister parietal bone. Some of the cranial/skull defect datasets are freely available online for download in the anonymized STL format. The datasets can be used for own research purposes, but we kindly ask to cite our work [20]. All relevant data are hosted at the public repository Figshare. Please see data hosted at Figshare at the following URL: https://figshare.com/articles/Cranial_Defect_Datasets/4659565

Note: when more datasets get available over time or other researchers provide us new cases, we will add them to the database. Furthermore, our data collection includes the software network (note that our software prototype is not a FDA/CE-certified application). Finally note that the original CT data can allow a finer review of a generated implant, e.g. checking the fitting, measuring the bone thickness for the fixation screws and checking the mirroring based on a self-defined midsagittal plane. However, if someone has own CT data, (s)he can also load it into the network and overlay it with the planned implant for a finer review of the results as mentioned before.

Workflow overview–We implemented a MeVisLab network in this contribution that uses manually set markers and a mirrored skull to generate a first curvature template for the defected area. Additionally, Delaunay-Triangulation [21] is used to construct a triangulated mesh via manually placed markers. Furthermore, we implemented a custom created smoother-module that rearranges the markers based on Laplacian-smoothing, which is a common and well-fitting method to smooth a generated mesh [22], [23]. This helps to avoid hard edges and generates a smooth and more aesthetic looking implant. The next sections describe the network and its modules in more detail.

Network–In Fig 2 the constructed network, created within MeVisLab, is shown. Each block represents a module with different functions. Beside already pre-defined modules, offering basic and advanced algorithms, the platform also provides an interface for the implementation of new modules using C++. The different colors represent different types of modules, where blue ones process WEM data, green modules mark an Open Inventor module, working with visual scene graphs (3D) and last, the brown ones, hiding sub networks formed by the other modules. To connect the modules with each other, again, two different connector types are available: The undirected lines, for a data connection and the directed ones, indicated by the arrow, for a parameter connection. When it comes to the inputs/outputs, one can distinguish between three types: The squared for Base-objects, pointing to data structures, the triangulated for ML images and the half-circled, for Inventor scenes. With the introduced modules and connections, the network was set up, consisting of four main parts:

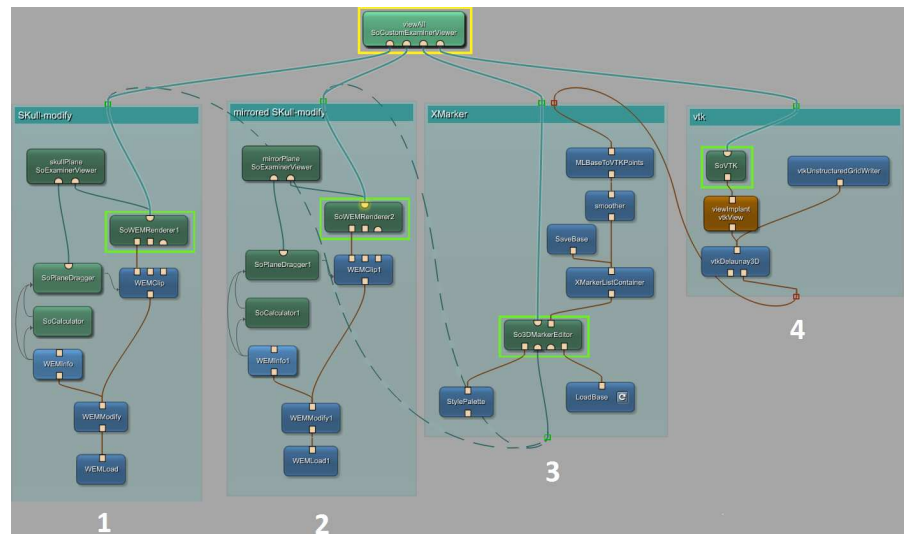

**Fig 2. MeVisLab network used for the implant generation.** The dashed lines indicate alternate usage depending on which skull the markers are placed (original or mirrored). Overall the network can be subdivided into four groups (1–4): loading and modifying the original skull (left), loading and modifying the mirrored skull including mirroring (second block from the left), performing marker operations like setting, smoothing, etc. (third block from the left) and the vtk environment/operations including Delaunay triangulation (rightmost).

doi:10.1371/journal.pone.0172694.g002

1. Skull modify–This section loads the dataset once and handles its preparation. The user can apply transformations in form of rotation, translation, applying a transform matrix and much more. Further, the skull can be cut with a plane, where just one side of this plane will be visible. This option is necessary to get a view on the inner parts of the skull, for example, the inner edges. The output is a modified skull.

2. Mirrored skull modify–This section holds exactly the same network components as the one described before. However, it has its purpose in preparing the second dataset, which is mirrored and thus serving as a template. An example is presented in Fig 3, which also shows that pure mirroring without further processing is not sufficient to generate an implant. The reason is, that human skulls are in general to asymmetric and therefore an implant resulting from pure mirroring has in most cases to be further adapted and refined. Further note that in general the exact midsagittal plane cannot be determined without the CT data. In our software an initial (mid)sagittal plane is defined via the nose and the eye sockets. Then, the user can make refinements by interactively reposition the plane and at the same time observing the outcome on the defected side. However, if an user has own cases with the CT data available, (s)he can overlay the data in our software and use this additional information to define the exact midsagittal plane.

3. XMarker–In this part all marker operations are handled, from setting the markers, storing them and performing modifications, until finally converting them from an (MeVisLab) MLBase type to vtkPoints.

4. vtk–This section uses only modules based on the vtk library, like the calculation of the Delaunay-Mesh with *vtkDelaunay3D* and has its purpose in generating the triangulated mesh and saving the data, respectively the final implant in STL format for 3D printing.

Implant Generation Pipeline–The workflow pipeline can be separated in three big parts, almost according to the four main network groups. Fig 4 gives an illustrative description, which is further described in the following sections.

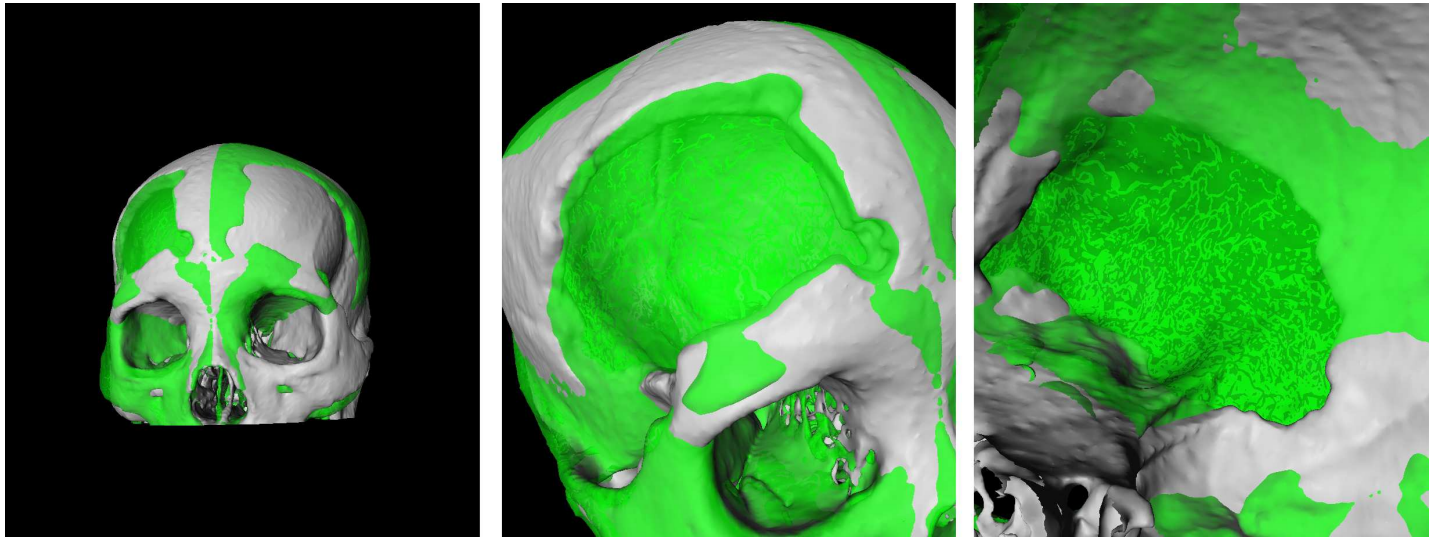

**Fig 3. Example of pure mirroring of a skull without further surface processing.** The left image shows a frontal view of the original skull (white) and the mirrored skull (green). The image in the middle shows a close-up view of the defected area. Finally, the right image shows a close-up view of the defected area from the inside.

## Dataset preparation: Original skull, mirrored skull (Fig 4, upper block)

First, by clicking on the left WEM load module in the skull modify section, the path to the file is chosen where the skull with the cranial defect is stored. Second, the WEM modify module is used for centring the data set in world coordinates. With the skull plane viewer module the user is able to cut the dataset with a plane, thus allowing to view the defect also from the inside. Since the data is used twice in the network, the dataset has also to be loaded in the mirrored skull modify section, where it serves as a template. Same actions need to be performed with this dataset and additionally in the WEM modify module a mirroring operation has to be performed by applying the appropriate transformation matrix. This matrix is a diagonal matrix where the first entry is inverted. Fig 5 shows the result of these preparation steps where the white part illustrates the original and cut dataset and the green part the mirrored skull.

## Marker setting (Fig 4, middle block)

The first step in this section is to set markers on the edge of the defect. Next, a connection from the first WEM renderer to the 3D marker editor has to be established, which will enable to set markers on the surface of the original skull. Overall, we have four marker types: outer edge, inner edge, outer surface and inner surface, which are differentiate by their marker types (values zero to three) in the marker editor options panel. In the *viewAll* viewer, it is now possible to set the markers along the outer edge of the defect. Next, the marker type has to be changed from zero to one and the inner edge markers on the original skull's surface are set (note: using the cut plane option from skull modify). Before being able to set the outer surface markers, a connection from second WEM renderer in the mirrored skull modify section to the 3D marker editor is established. Furthermore, it is necessary to connect the mirrored skull to the view all viewer. If the marker type is set to two, the network is ready for setting markers on the mirrored surface. However, further improvements can be achieved by using the translation operations in the WEM modify module or by setting markers just in regions where the surface fits already perfectly. Then adjusting the mirrored skull by translating it again, so that other regions fit perfectly, and set additional markers on the surface. The final step it to set the inner surface markers, the same way as it was done with the outer surface markers but instead using the marker type three.

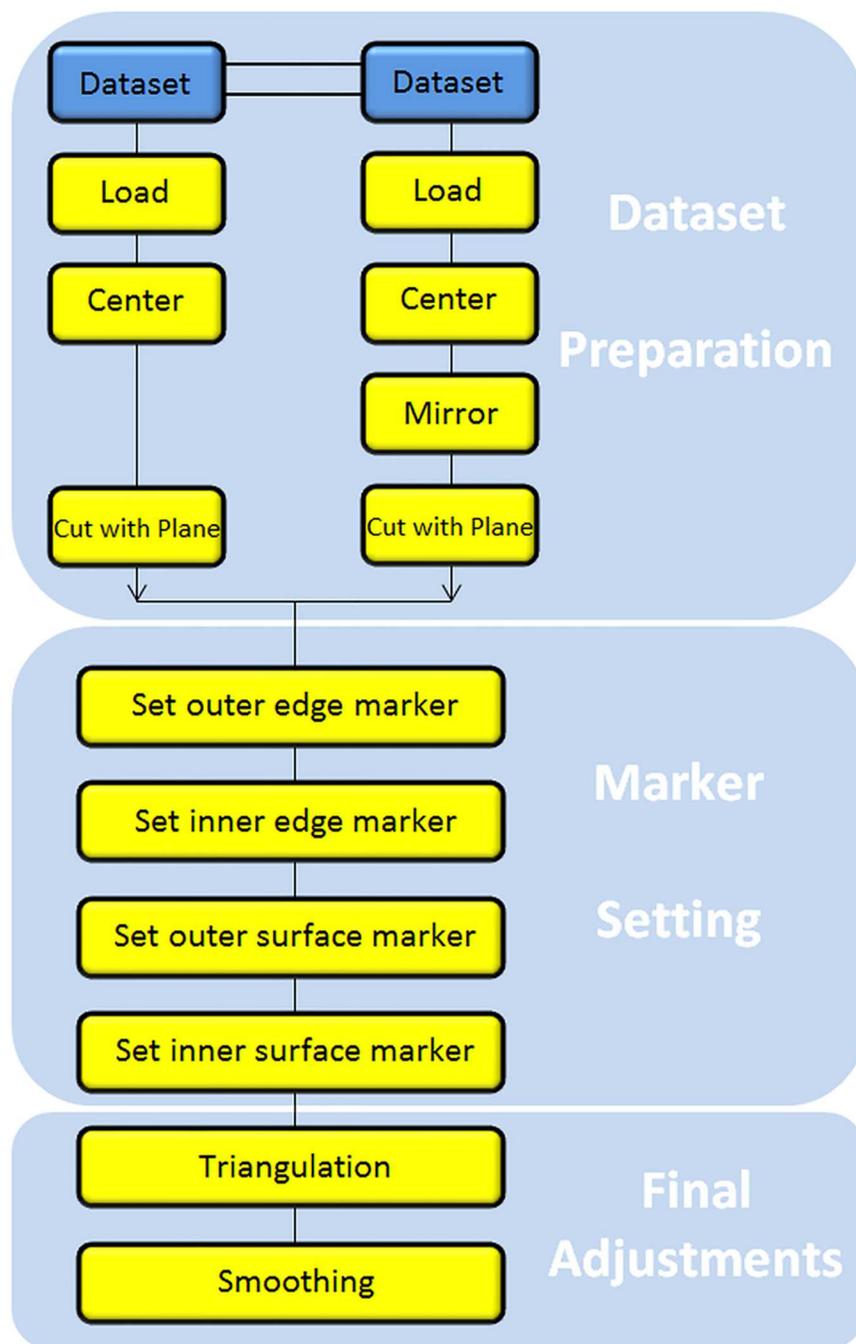

**Fig 4. The overall workflow pipeline separated in three main parts: Dataset preparation, marker setting and final adjustments.**

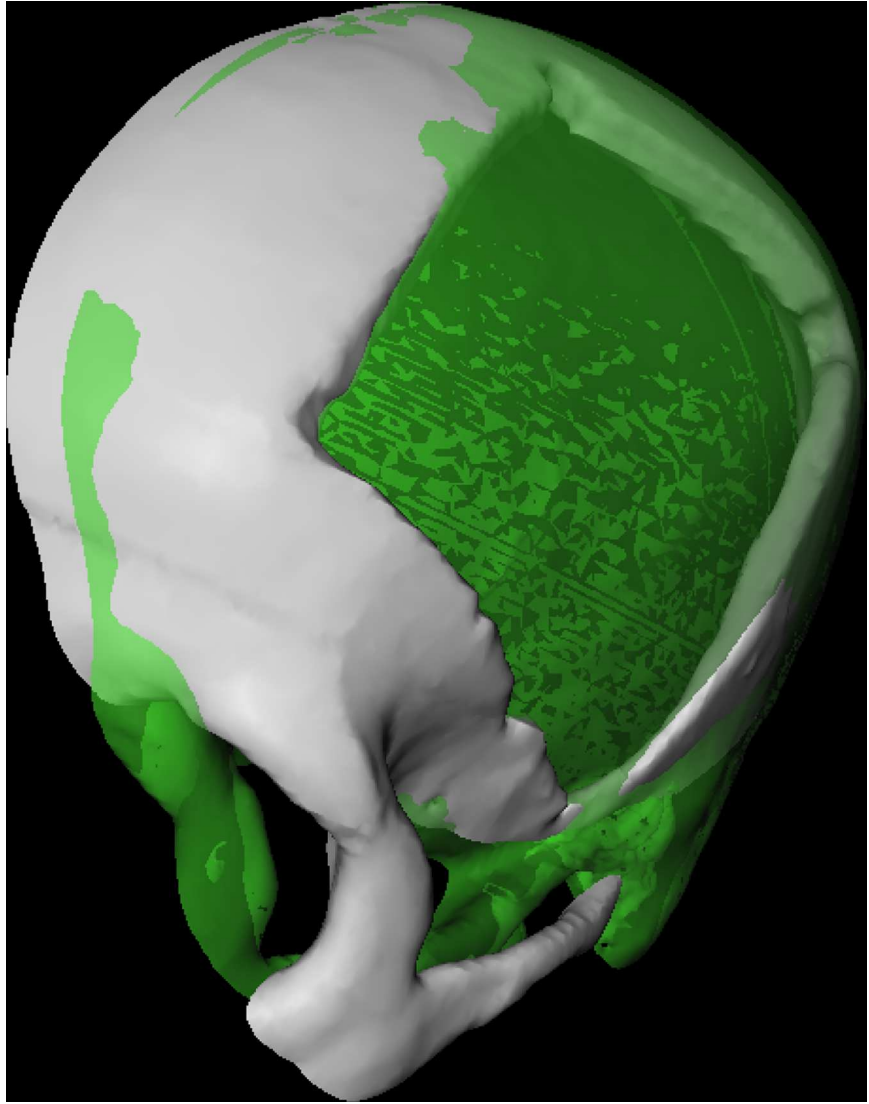

**Fig 5. Result of the preparation steps with the original skull in white and the mirrored skull in green.** Both skulls are cut with a plane, to have a view on the inner as well the outer surface.

## Final adjustments save/load (Fig 4, lower block)

After the triangulation has been performed, adjustments can be applied to gain an even better looking implant. By clicking/opening the smoother module, the radius (x) as well as the weight of the edge markers (y) can be adjusted to improve the implant by smoothing it. In addition, the characteristics of the vtk Delaunay 3D module can be adjusted as well, to improve the outcome by changing its parameters. Finally, the marker list can be saved and loaded at any time, as well as the generated implant in the STL file format.

Smoother Module–Since the smoother module was a new implementation, the following lines together with the flow chart in Fig 6 give a rough description of the module and how it works and how it was set up. First, a ML module was set up using one input, one output and two field variables serving as user input. The header file of the module's C++ data is set up in a standard way (except the inclusion of the vector library), consisting of a constructor, a method for handling the changes of the fields and the definitions of the input and output marker lists, using base field pointers. The parameters x and y are set as integer field pointer variables and a flag that shows if parameters have changed, and an internal storage of a marker-list are also used. Having a look at the source file, we first implemented the constructor, by giving it zero inputs and zero outputs from MeVisLab's point of view, which is actually not true, since we handle the inputs and outputs with the base field pointers, declared before. During setting up the constructor, the handle notification method is disabled. More, the field parameters are set up, as well as the base outputs for the marker lists. In the implementation of the handle notification method, we first check for the parameters value to be more than zero or one. Next, it is checked if the parameters have changed. If so, the output list is cleared and the input is checked for validity. Further, the operation is dismissed if the list consists of less than three markers. Otherwise, the input list is copied to the internal marker list to start the actual algorithm, which can be observed in the code snippet in Fig 7. In more detail, the code snippet begins with iterating over all nodes (markers). If the current node is an edge marker, nothing is done, since those are already fixed in position (1). Otherwise, it is iterated over all nodes again to calculate the distance between the current node and all other nodes (2). Therefore, they have to be from the same layer (inner and outer) and if so, the distance is calculated. If furthermore the distance criterion, defined by the y parameter, is fulfilled, the node is saved into a vector. After those iterations, a vector is accessible, which holds all the nodes used for the Laplacian smoothing of the current node. Now, the new position is calculated (3) and (4) based on the nodes stored in the vector as seen in following equation:

$$\bar{x}_i = \frac{1}{N} \sum_{j=1}^{N} \bar{x}_j \tag{1}$$

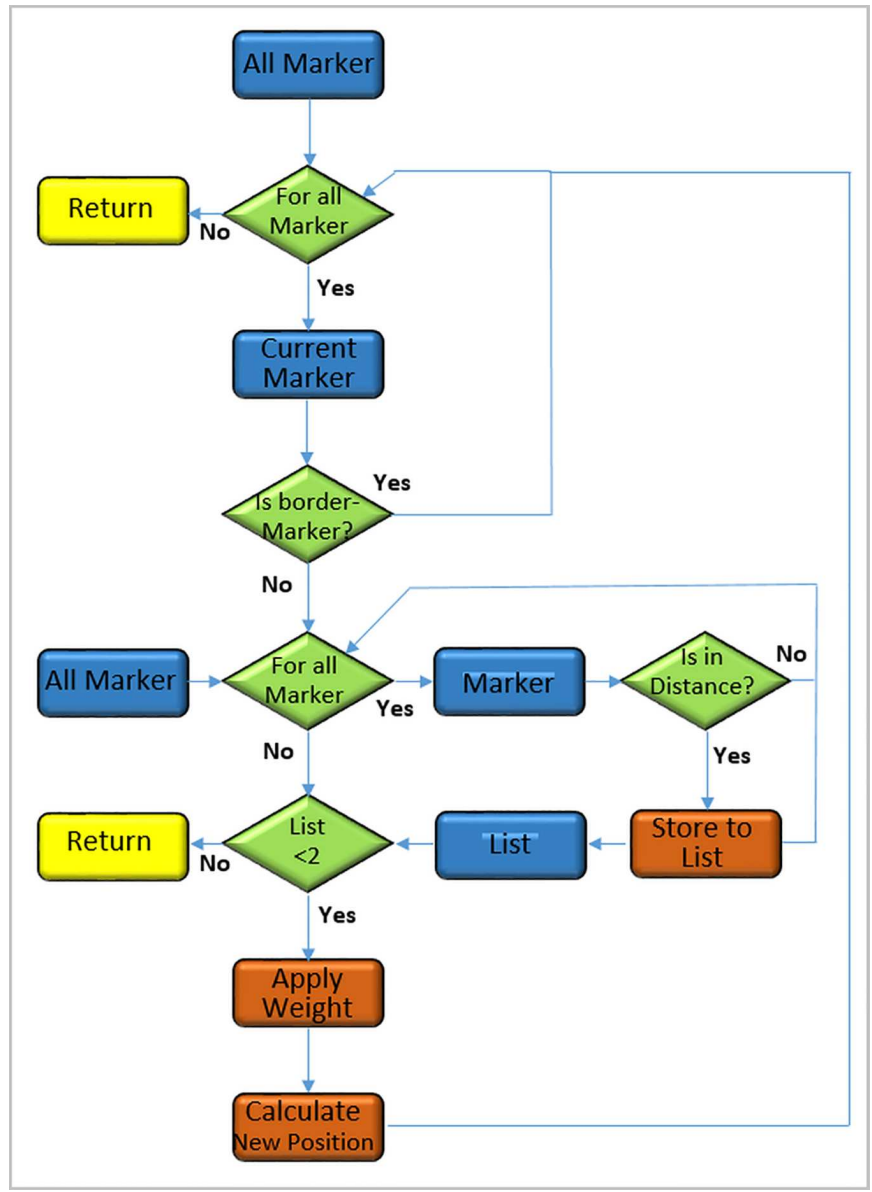

**Fig 6. Overall flowchart of the smoothing algorithm, which is based on an own C++ MeVisLab module.** Blue boxes represent object types, green diamonds are test expression and orange boxes belong to processing steps.

Whereby, $N$ is the number of adjacent nodes to node $i$ in the given radius, set by the $x$ parameter, $x_j$ is the position of the $j^{th}$ adjacent node, and $x_i$ is the new position for node $i$. Improvements made on the basic Laplacian smoother are: (1.) The additional usage of the $x$-parameter to use just markers in the defined radius rather than all markers in the list. (2.) The so-called border markers are set as "true" markers, not being able to be moved in this calculation, since they are already set on the border by the user. Moreover, by applying a weight to these "true" markers, they gain more influence in the calculations, thus improving the final implant result even more. Note that smoothing can also cause shrinkage of an implant. However, the border markers ensure that the implant is always connected to the bone (if placed there by the user).

Evaluation–Our software has been evaluated with several clinical experts from the Department of Neurosurgery at the Medical University of Graz in Austria. All physicians know the state of the art in cranial implant planning and are familiar with commercial and non-commercial software solutions. However, our evaluation consisted of three main steps. In the first step, the so-called training phase, the developer of the software showed the physicians the planning of an implant from start to finish. In the next step, the physicians have been asked to perform the planning of an implant on their own, beginning with loading a new medical dataset. In the final step, the physicians were asked to fill out a questionnaire to rate the overall software solution. Generally, the questionnaire evaluates the software ergonomic formulated in general terms without reference to situations of use, application, environment or technology [24]. However, questions like "The software does not need a lot of training time" are seen in context of the current user and his/her experience/knowledge in cranial implant planning (in respect to other software solutions). Furthermore, the time was measured during the implant planning. As a feedback summary from the clinical partners, skull mirroring, marker setting and the smoothing was integrated into a user-friendly planning prototype. However, this is also a result of the hand in hand software development between the technical and clinical partners, were constant feedback from regular meetings had been taken into consideration and constantly been integrated into the software. For the questionnaire, the following questions had been asked (derived from ISONORM 9241/10):

1. The software does not need a lot of training time

2. The software is adjusted well to achieve a satisfying result

3. The software provides all necessary functions to achieve the goal

```
//calculate for all nodes
for (MLusize_t j = 0; j<nbrXMarkers; j++) {
    //if edge Marker, do not smooth
    if ((_outputXMarkerList[j].type == 0)         ┌─────┐
    || (_outputXMarkerList[j].type == 1)){}       │  1  │
    else{                                          └─────┘
        //calculate distance for all nodes         ┌─────┐
        nodes.clear();                             │     │
        for (MLusize_t k = 0; k < nbrXMarkers ; k++){
            // checks similarity of layer          │     │
            if (!((_outputXMarkerList[k].type      │     │
                + _outputXMarkerList[j].type) % 2)) {│    │
            dist_vec = _outputXMarkerList[j].pos    │  2  │
            - outputXMarkerList[k].pos;             │     │
            dist = sqrt((dist_vec[0] * dist_vec[0]) │     │
                    + (dist_vec[1] * dist_vec[1])   │     │
                    + (dist_vec[2] * dist_vec[2])); │     │
            //if distance criterion is fullfilled,  │     │
            //save node in vector                   │     │
            if ((dist < 0) && (dist <= _xFld->getIntValue()))
                nodes.push_back(k);                 └─────┘
            }
        }
    }
    if (nodes.size() > 1){
        sum = sum - sum; //reset sum
        //first value is the position of current node
        sum = _outputXMarkerList[j].pos;
        //counter for extra nodes, used in weight function
        nodes_plus = 0;
        for (int n = 1; n <= nodes.size(); n++){    ┌─────┐
            //if the node in the list is an edge node,│    │
            //apply weight                          │     │
            if ((_outputXMarkerList[nodes[n - 1]].type == 0)
            || (_outputXMarkerList[nodes[n - 1]].type == 1)){│
                sum = sum                           │     │
                +(_outputXMarkerList[nodes[n - 1]].pos│    │
                *((MLdouble)_yFld->getIntValue()));  │  3  │
                nodes_plus = nodes_plus             │     │
                    + (int)_yFld->getIntValue()-1;   │     │
                //-1 because one time is already     │     │
                //included for current node.         │     │
            }                                        │     │
            else{                                    │     │
                sum = sum                            │     │
                + _outputXMarkerList[nodes[n - 1]].pos;│   │
            }                                        └─────┘
        }

        outputXMarkerList[j].pos = sum              ┌─────┐
            / ((MLdouble) nodes.size() +1 +nodes_plus);│ 4 │
        //+1 for current node                       └─────┘
    }
}
.
.
.
```

**Fig 7.** Extract of the C++ code of the smoothing algorithm divided into four parts: checking for border markers (1), distance calculation (2), summing up all nodes used for calculating new position (3) and Calculating new position (4).

4. The software is not complicated to use

5. How satisfied are you with the UI surface? (arrangement, style, clarity)

6. How accurate was the placement of the implant?

7. How satisfied are you with the presented result?

8. Was it easy to adjust the implant? (position, orientation, model,...)

9. How satisfied have you been with the time consumed? (no training)

10. Overall impression

11. Would you use the software in a daily routine?
    (Assumed that the 3D printed implant would fit with just a few adjustments)

12. Time used for gaining satisfying result

## Results

Table 1 and Fig 8 presents the results of the questionnaire which were rated using a six-point Likert scale with increasing consent from one to six [25]. Fig 9 demonstrates the stepwise workflow during the implant generation process for a clinical case with a cranial defect on the left side. The left image shows the original skull (white) overlapped with the mirrored one (green). The figure in the middle shows the border markers (green) and surface markers (magenta). The rightmost image shows the smoothed implant (green). Fig 10 presents a side-by-side comparison of an unsmoothed (left) and smoothed (right) implant (green) for a clinical case with a large cranial defect on the left side. The improvement of the implant surface is indicated by red arrows and the bottom images show the inner view. Fig 11 shows implant planning results (green) of a clinical case with a large and complex cranial defect on the right side of the frontal area. The unsmoothed implant is shown in the lower left image and a well-fitting (no gaps between the implant and the bone) and aesthetic looking result after additional Laplacian smoothing is shown in the lower right image (indicated by red arrows, respectively). Moreover, Figs 12 and 13 present mirroring and implant planning results (green) of a clinical case with a large cranial defect on the left side. Furthermore, a planned implant can be 3D printed via its STL file, as seen in the photo of Fig 14 (yellow). In addition, the photo shows the corresponding patient skull which has also been 3D printed for testing purposes (white). The planning time for cranial 3D implant could be reduced to about thirty minutes for a new medical user. However, for a user familiar with the software, the planning time could even be reduced to about fifteen minutes. This stands in strong contrast to a reported planning time of several hours for Blender, where our clinical partners shift vertex per vertex of a geometrical form into position and traditional softwares (such as Imageware, UG, and Magics RP) with a planning time around two hours. The overall planning starts with loading the STL file into our software (as stated before the transformation from CT/DICOM data to STL can be done with the MeVisLab *WEMIsoSurface* module or Slicer [26] in a few minutes). Typical times for an

**Table 1. Result of evaluation questionnaire.** Q1-Q11 according questions and T for the used time in minutes. Rated with a six-point Likert scale with increasing accordance from one to six.

| | Subject | | | | |
|---|---|---|---|---|---|
| Nr. | 1 | 2 | 3 | Median | error |
| Q1 | 5 | 5 | 6 | 5.33 | 0.27 |
| Q2 | 5 | 6 | 6 | 5.67 | 0.27 |
| Q3 | 4 | 6 | 6 | 5.33 | 0.54 |
| Q4 | 4 | 6 | 6 | 5.33 | 0.54 |
| Q5 | 5 | 5 | 5 | 5.00 | 0 |
| Q6 | 3 | 4 | 6 | 4.33 | 0.72 |
| Q7 | 3 | 6 | 6 | 5.00 | 0.81 |
| Q8 | 3 | 6 | 6 | 5.00 | 0.81 |
| Q9 | 4 | 5 | 6 | 5.00 | 0.47 |
| Q10 | 4 | 5 | 6 | 5.00 | 0.47 |
| Q11 | 6 | 6 | 6 | 6.00 | 0 |
| T | 30 | 20 | 4 | 17.83 | 6.07 |

doi:10.1371/journal.pone.0172694.t001

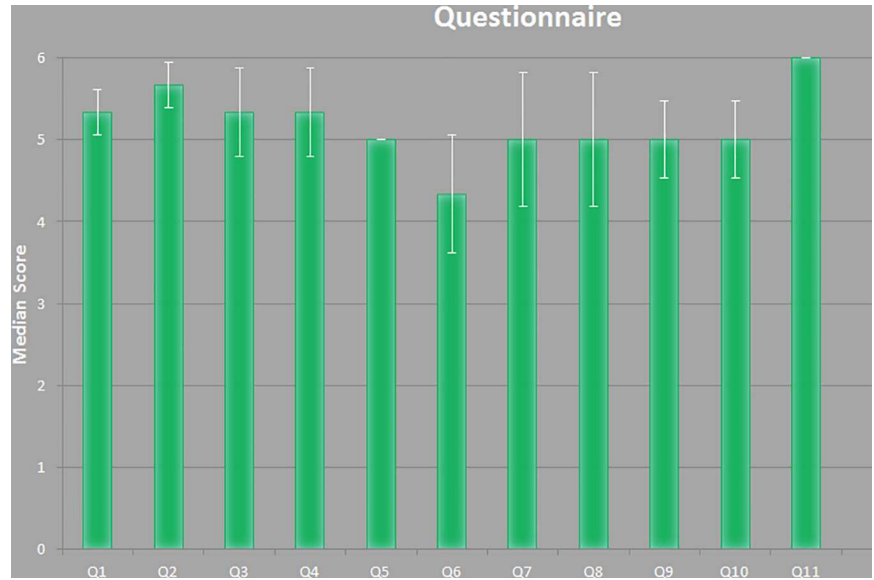

**Fig 8. Results of the questionnaire which were rated using a six-point Likert scale with increasing consent from one to six.**

experienced user for all workflow steps until the final implant is generated, are as follows (for the colors please refer to the images):

0. Starting the software and loading the network (0:00)

1. Load dataset (0:25)

2. Load dataset again for mirroring (0:33)

3. View original skull in white (0:48)

4. View skull for mirroring in green (1:15)

5. Register green skull to white (1:21)

6. Mirror green skull (1:32)

7. Set clipping plane for original skull (1:44)

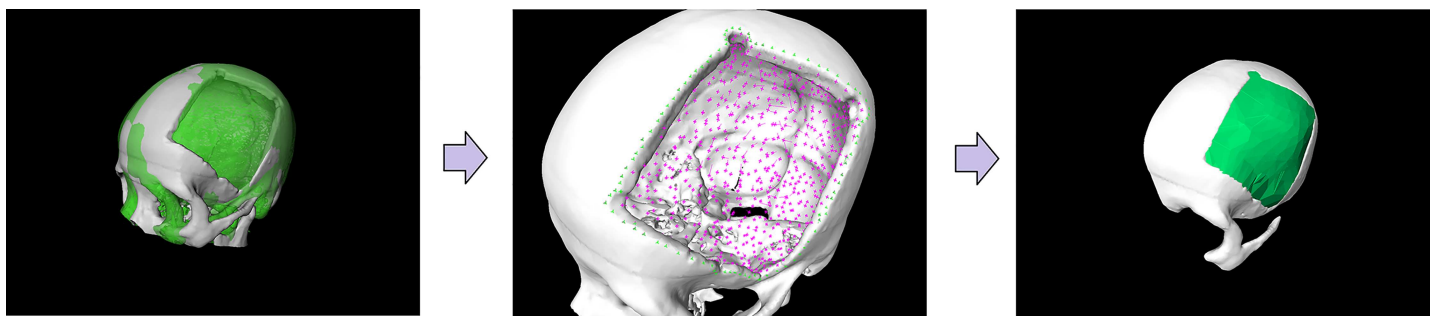

**Fig 9. Stepwise workflow during the implant generation process for a clinical case with a cranial defect on the left side.** The left image shows the original skull (white) overlapped with the mirrored one (green). The figure in the middle shows the border markers (green) and surface markers (magenta). The rightmost image shows the smoothed implant (green).

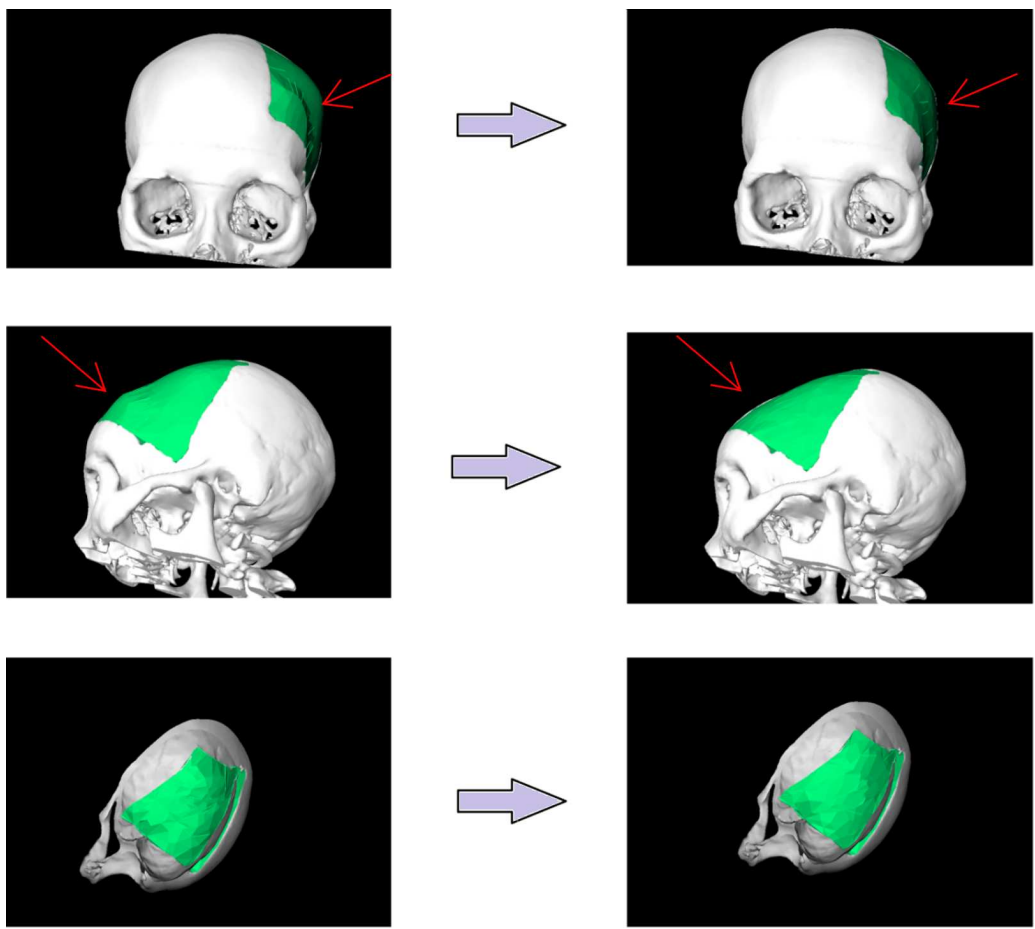

**Fig 10. Side by side comparison of an unsmoothed (left) and smoothed (right) implant (green) for a clinical case with a large cranial defect on the left side.** The improvement of the implant surface is indicated by red arrows and the bottom images show the inner view.

doi:10.1371/journal.pone.0172694.g010

8. Set clipping plane for mirrored skull (2:12)

9. Set markers on the outside edge of the defect (2:40)

10. Set markers on the inside edge of the defect (3:13)

11. Set markers on the outside surface (3:59)

12. Set markers on the inside surface (4:44)

13. Perform initial Delaunay triangulation (5:55)

14. Adapt/refine Delaunay triangulation (6:19)

15. Perform Laplacian-smoothing on the implant (7:29)

16. Inspection of the result, further refinement and saving the implant

Finally, for an inter-user accuracy analysis, we compared two planned implants from two users via the Dice Similarity Coefficient (DSC) [27] and the Hausdorff Distance (HD) [28].

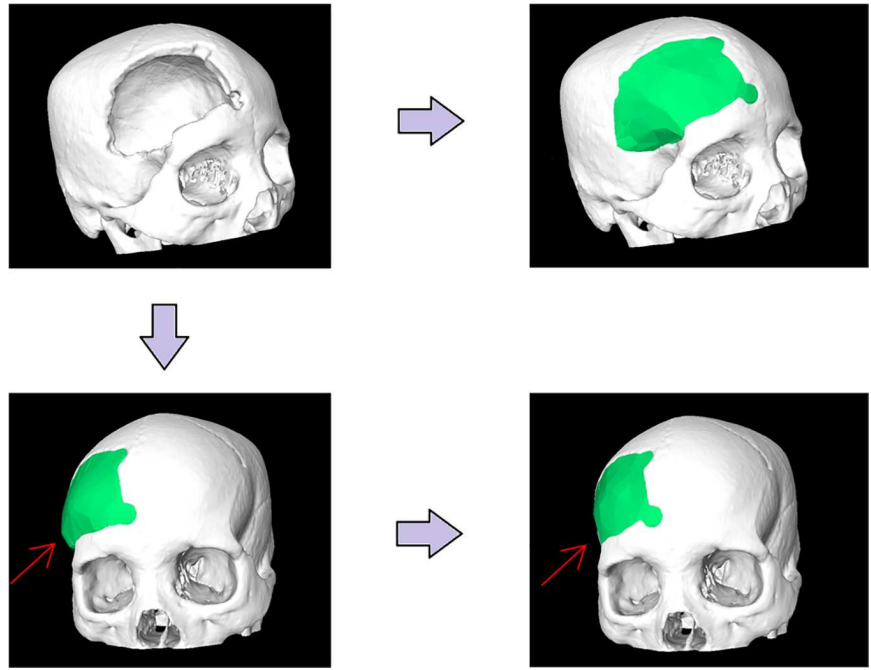

**Fig 11. Implant planning results (green) of a clinical case with a large and complex cranial defect on the right side of the frontal area.** The unsmoothed implant is shown in the lower left image and a well-fitting and aesthetic looking result after additional Laplacian smoothing is shown in the lower right image (indicated by red arrows, respectively).

The results yielded to a DSC of 88.08% and a HD of 5.41 voxel. This shows how well two implants agree when planned with our software even between users:

User 1: 28581775 voxel, 115900 mm$^3$
User 2: 28450794 voxel, 115369 mm$^3$

## Discussion

The constructing of cranial 3D implants is a challenging task and lacks on free software especially designed for the specific planning tasks. Commercially available software, like Mimics/

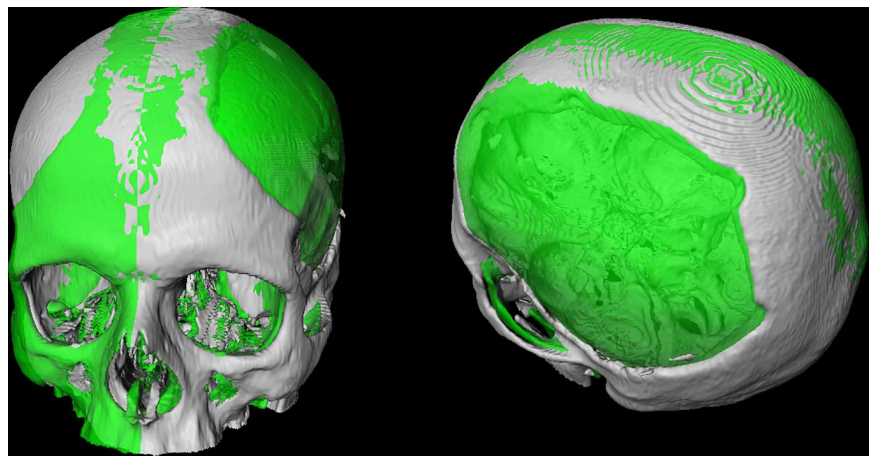

**Fig 12. Pure mirroring results (green) of a clinical case with a large cranial defect on the left side of a patient's skull.** Frontal (left) and upper (right) views.

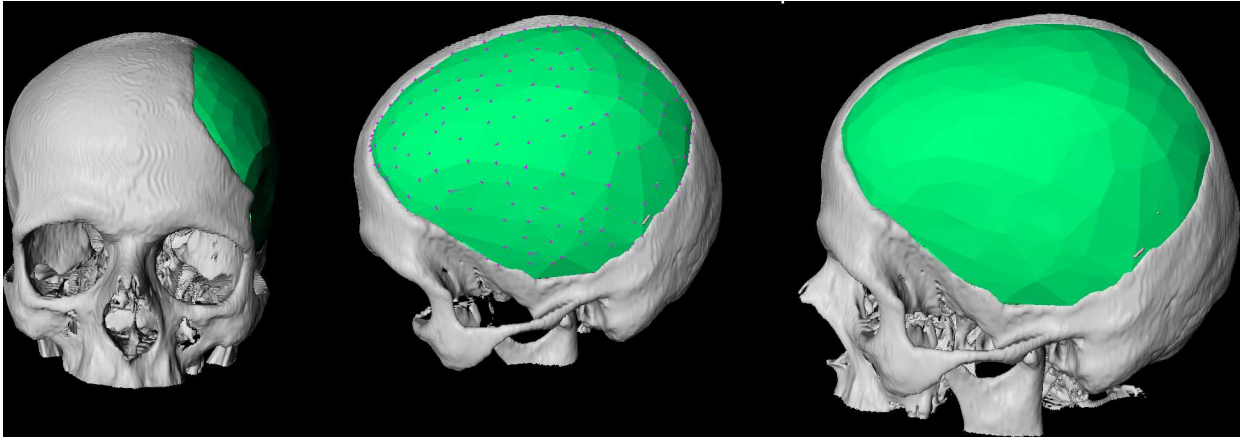

**Fig 13. Implant planning results (green) of a clinical case with a large cranial defect on the left side from Fig 11.** Frontal (left) and side (middle and right) views. In addition, the surface markers (magenta) are shown in the middle image.

3-matic, KLS Martin or companies that provide services for these applications, can be very complex and expensive, and may also not be available in a clinic. In this contribution, however, the semi-automatic planning of cranial 3D Implants has been demonstrated under the medical prototyping platform MeVisLab, which is free for own research purposes. The MeVisLab prototype consists of a customized data-flow network and the results showed that MeVisLab can be an alternative to complex commercial planning software. Overall, the user is guided through the workflow of the software prototype, starting with loading and mirroring the patients head, to generate an initial curvature of the cranial implant. However, because human heads are in general too asymmetric, pure mirroring is not sufficient to generate an implant and our software enables Delaunay triangulation with Laplacian smoothing to generate an aesthetic looking and well-fitting implant. Moreover, our software prototype allows the user to save the designed 3D model of the cranial implant as a STL file for 3D printing. The printed cranial implant can be used for further pre-interventional planning, like verifying the size and shape (Fig 12), or it can even be used as the real 3D implant for the patient in an following operation. In summary, the achieved highlights of this contribution are:

- The successful interactive planning and reconstruction of cranial 3D implants under MeVisLab;

- Enables skull mirroring to obtain an initial curvature of the implant for the defect restoration;

- Offers additional smoothing and triangulation for an aesthetic looking and well-fitting outcome;

- Has been evaluated with real patient Computed Tomography (CT) data from the clinical routine;

- Allows 3D implant exporting as Computer-Aided Design (CAD) file format for 3D printing and in-depth pre-surgical assessment;

- Providing the source code and clinical datasets to the research community for own usage.

There are several areas of future work, including an evaluation with a greater patient collective as we collect more clinical datasets, and intra- and inter-rater reliability ratings of the planned implants. Furthermore, a comprehensive comparison and evaluation study with commercial available software products. Therefore, collecting datasets and real manufactured implants from the clinical routine, which will be 3D scanned to compare them with the planned implants from our software. Moreover, a direct comparison with an open-source software developed by colleagues at the Jiao Tong University in China [29]. Finally, testing the software in areas where 3D printing technology plays also an important role, like oral implantology, pelvis and cervical vertebrae implants [30] and integration into a medical AR navigation system [31].

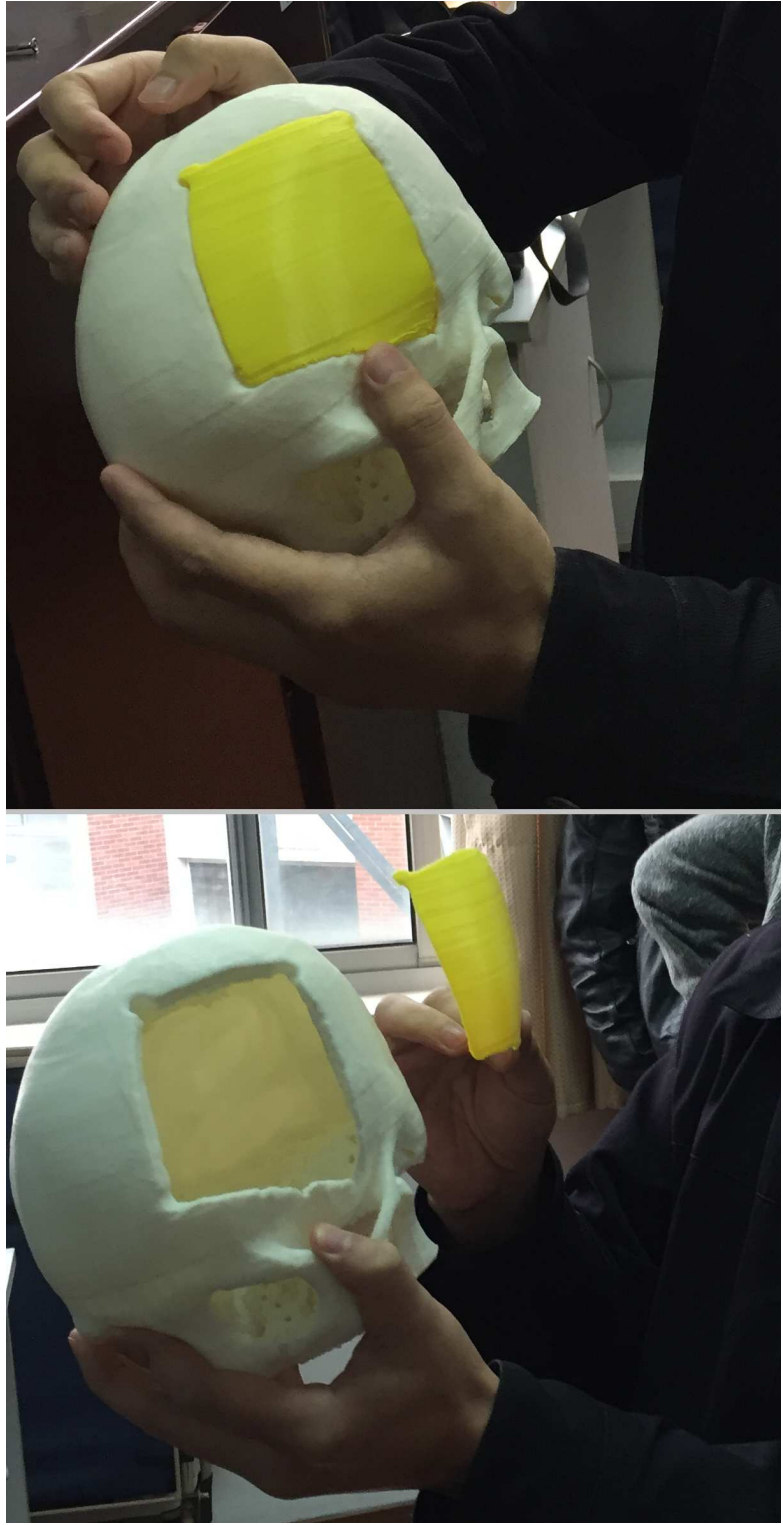

**Fig 14. The photo shows the inspection of a 3D printed implant (yellow) and the corresponding 3D printed patient skull (white), which has a large defect on the left side.**

## Acknowledgments

The work received funding from BioTechMed-Graz in Austria (Hardware accelerated intelligent medical imaging) and the 6[th] Call of the Initial Funding Program from the Research & Technology House (F&T-Haus) at the Graz University of Technology (PI: Jan Egger). Dr. Xiaojun Chen receives support by the Natural Science Foundation of China (Grant No.: 81511130089) and the Foundation of Science and Technology Commission of Shanghai Municipality (Grants No.: 14441901002, 15510722200 and 16441908400). A tutorial video demonstrating the cranial implant planning be found under the YouTube channel: https://www.youtube.com/c/JanEgger/videos. Cranial/skull defect datasets are freely available online for own research studies (please cite this paper if you use these in your work). All relevant data are hosted at the public repository Figshare. Please see data hosted at Figshare at the following URL: https://figshare.com/articles/Cranial_Defect_Datasets/4659565. Moreover, our data collection includes the software network from this contribution.

## Author Contributions

**Conceptualization:** MG JE.

**Data curation:** MG AT MÜ UZ XL GvC US DS XC JE.

**Formal analysis:** MG AT MÜ GvC JE.

**Funding acquisition:** DS XC JE.

**Investigation:** MG AT MÜ GvC.

**Methodology:** MG JE.

**Project administration:** DS JE.

**Software:** MG JE.

**Supervision:** JE.

**Validation:** MG AT MÜ GvC.

**Visualization:** MG JE.

**Writing – original draft:** MG JE.

**Writing – review & editing:** MG JE.

## References

1. Gordon CR, Fisher M, Liauw J, Lina I, Puvanesarajah V, Susarla S. et al. Multidisciplinary Approach for Improved Outcomes in Secondary Cranial Reconstruction: Introducing the Pericranial-onlay Cranioplasty Technique. Neurosurgery 2014; 10(2): 179–189.

2. Dujovny M, Aviles A, Agner C, Fernandez P, Charbel FT. Cranioplasty: cosmetic or therapeutic?. Surg Neurol. 1997; 47(3): 238–41. PMID: 9068693

3. Mangubat E, Sani S. Acute global ischemic stroke after cranioplasty: case report and review of the literature. Neurologist. 2015; 19(5): 135–9. doi: 10.1097/NRL.0000000000000024 PMID: 25970836

4. Brommel T, Rydning PN, Pripp AH, Helseth E. Cranioplasty complications and risk factors associated with bone flap resorption. Scandinavian Journal of Trauma, Resuscitation and Emergency Medicine 2015; 23(1): 1–7.

5. Replogle RE, Lanzino G, Francel P, Henson S, Lin K, Jane JA. Acrylic cranioplasty using miniplate struts. Neurosurgery 1996; 39(4): 747–749. PMID: 8880768

6. Itokawa H, Hiraide T, Moriya M, Fujimoto M, Nagashima G, Suzuki R. et al. A 12 month in vivo study on the response of bone to a hydroxyapatite-polymethylmethacrylate cranioplasty composite. Biomaterials 2007; 28(33): 4922–4927. doi: 10.1016/j.biomaterials.2007.08.001 PMID: 17707904

7. Sprio S, Fricia M, Maddalena GF, Nataloni A, Tampieri A. Osteointegration in cranial bone reconstruction: a goal to achieve. J Appl Biomater Funct Mater. 2016; 14(4):e470–e476. doi: 10.5301/jabfm.5000293 PMID: 27311430

8. Parthasarathy J. 3D modeling, custom implants and its future perspectives in craniofacial surgery. Annals of Maxillofacial Surgery. 2014; 4(1): 9–18. doi: 10.4103/2231-0746.133065 PMID: 24987592

9. Lee MY, Chang C-C, Lin C-C, Lo L-J, Chen Y-R. Custom implant design for patients with cranial defects. Engineering in Medicine and Biology Magazine 2003; 21(2): 38–44.

10. van der Meer WJ, Bos RR, Vissink A, Visser A. Digital planning of cranial implants. Br J Oral Maxillofac Surg. 2013; 51(5): 450–452. doi: 10.1016/j.bjoms.2012.11.012 PMID: 23266152

11. Chulvi V, Cebrian-Tarrasón D, Sancho Á, Vidal R. Automated design of customized implants. Revista Facultad de Ingeniería Universidad de Antioquia 2013; 68: 95–103.

12. Egger J, Kapur T, Fedorov A, Pieper S, Miller JV, Veeraraghavan H, et al. GBM Volumetry using the 3D Slicer Medical Image Computing Platform. Sci Rep. 2013; 3: 1364. doi: 10.1038/srep01364 PMID: 23455483

13. Lo JL, Chen YR, Tseng CS, Lee MY. Computer-aided reconstruction of traumatic fronto-orbital osseous defects: aesthetic considerations. Chang Gung Medical Journal.2004; 27(4): 283–291. PMID: 15239195

14. Marreiros FM, Heuzé Y, Verius M, Unterhofer C, Freysinger W, Recheis W. Custom implant design for large cranial defects. Int J Comput Assist Radiol Surg. 2016; 11(12): 2217–2230. doi: 10.1007/s11548-016-1454-8 PMID: 27358081

15. Yu W, Lia M, Lib X. Fragmented skull modeling using heat kernels. Graphical Models. 2012; 74(4): 140–151.

16. Li X, Yin Z, Wei L, Wan S, Yu W, Li M. Symmetry and template guided completion of damaged skulls. Computers and Graphics 2011; 35(4): 885–893.

17. Gilardino MS, Karunanayake M, Al-Humsi T, Izadpanah A, Al-Ajmi H, Marcoux J, et al. A comparison and cost analysis of cranioplasty techniques: autologous bone versus custom computer-generated implants. Journal of Craniofacial Surgery 2015; 1: 113–117.

18. Egger J, Tokuda J, Chauvin L, Freisleben B, Nimsky C, Kapur T, et al. Integration of the OpenIGTLink network protocol for image-guided therapy with the medical platform MeVisLab. Int J Med Robot. 2012; 8(3): 282–90. doi: 10.1002/rcs.1415 PMID: 22374845

19. Tam MD, Laycock SD, Bell D, Chojnowski A. 3-D printout of a DICOM file to aid surgical planning in a 6 year old patient with a large scapular osteochondroma complicating congenital diaphyseal aclasia. Journal of Radiology Case Report 2012; 6(1): 31–37.

20. Gall M, Li X, Chen X, Schmalstieg D, Egger J. Cranial Defect Datasets. ResearchGate 2016.

21. Gopi M, Krishnan S, Silva CT. Surface reconstruction based on lower dimensional localized Delaunay triangulation. Computer Graphics Forum 2000; 19(3): 467–478.

22. Amenta N, Bern M, Eppstein D. Optimal point placement for mesh smoothing. Journal of Algorithms 1999; 30(2): 302–322.

23. Field DA. Laplacian smoothing and delaunay triangulations. International Journal for Numerical Methods in Biomedical Engineering 1988; 4(6): 709–712.

24. Prümper J. Der Benutzungsfragebogen ISONORM 9241/10: Ergebnisse Zur Reliabilität und Validität. In: R. Liskowsky u.a. (Hg.): Software-Ergonomie '97, Stuttgart; 1997.

25. Allen IE, Seaman CA. Likert Scales and Data Analyses. Quality Progress. 2007; 40(7): 64–65.

26. Egger J, Kapur T, Nimsky C, Kikinis R. Pituitary adenoma volumetry with 3D Slicer. PLoS ONE 7(12): e51788. doi: 10.1371/journal.pone.0051788 PMID: 23240062

27. Sampat MP, Wang Z, Markey MK, Whitman GJ, Stephens TW, Bovik AC. Measuring intra- and inter-observerser agreement in identifying and localizing structures in medical images. IEEE International Conference on Image Processing 2006; pp. 1–4.

28. Hausdorff F. Grundzuege der Mengenlehre. Veit & Comp., Leipzig 1914 (rep. in Srishti D. Chatterji et al. (Hrsg.), Gesammelte Werke, Band II, Springer, Berlin; 2002.

29. Li X, Xu L, Zhu Y, Egger J, Chen X. A semi-automatic implant design method for cranial defect restoration. Int J CARS  11 2016; (Suppl 1): S241–S243.

30. Chen X, Xu L, Yang Y, Egger J. A semi-automatic computer-aided method for surgical template design. Sci Rep. 2016;  6: 20280. doi: 10.1038/srep20280 PMID: 26843434

31. Chen X, Xu L, Wang Y, Wang H, Wang F, Zeng X, et al. Development of a surgical navigation system based on augmented reality using an optical see-through head-mounted display. J Biomed Inform. 2015;  55: 124–31. doi: 10.1016/j.jbi.2015.04.003 PMID: 25882923